\documentclass[12pt,paper]{JHEP3}
\usepackage{epsfig,graphicx,bm,amsmath}

\newcommand{\mc}[1]{\mathcal{#1}}

\title{Quasilocal equilibrium condition for black ring}

\author{Dumitru Astefanesei, Maria J. Rodriguez, and Stefan Theisen\\
Max-Planck-Institut f\"ur Gravitationsphysik,
Albert-Einstein-Institut, 14476 Golm, Germany\\

\\E-mail: \email{dumitru@aei.mpg.de, maria.rodriguez@aei.mpg.de, stefan.theisen@aei.mpg.de }}

\abstract{
We use the conservation of the {\it renormalized} boundary stress-energy
tensor to obtain the equilibrium condition
for a {\it general} (thin or fat) black ring solution. We also investigate
the role of the spatial stress in the thermodynamics of
deformation within the quasilocal formalism of Brown and York 
and discuss the relation with other methods. In particular, we discuss the quantum
statistical relation for the unbalanced black ring solution.}


\preprint{AEI-2009-085}

\begin{document}
\vfill \setcounter{page}{0} \setcounter{footnote}{0}
\setcounter{tocdepth}{2}
%%%%%%%%%%%%%%%%%%%%%%%%%%%%%%%%%%%%%%%%%%%%%%%%%%%%%%%%%%%%%%%%%%%%%%
\section{Introduction}
%%%%%%%%%%%%%%%%%%%%%%%%%%%%%%%%%%%%%%%%%%%%%%%%%%%%%%%%%%%%%%%%%%%%%%

A basic known fact of general relativity is that there is no concept of
local energy for the gravitational field. Due to the equivalence principle
it is possible to eliminate every observable effect of the gravitational
field in a suitable spacetime neighborhood.\footnote{Even if the gravitational
field can be measured by the geodesic deviation of two observers, a single
observer can not distinguish it from kinematical effects.} Therefore, in
general relativity the gravitational energy and momentum have non-local validity.

One of the most powerful frameworks for computing conserved quantities
in general relativity is the `quasilocal' formalism of Brown and York
\cite{Brown:1992br}. The basic idea in \cite{Brown:1992br} is to define
a `quasilocal' energy inside a given finite region (rather than defining
the energy at a point). Thus, an appealing feature of quasilocal energy
is its direct derivation from the gravitational action for a spatially
bounded region.

In this paper, we consider the quasilocal energy as applied to spacetimes
that are asymptotically flat in spacelike directions. The spatial
infinity --- the part of infinity which is reached along
spacelike geodesics --- is represented by one point in the Penrose diagram of
conformal compactification for Minkowski space. In this context, it is
better to visualize it as the hyperboloid of spacelike
directions (it is isometric to the unit $4$-dimensional de Sitter
space) --- a detailed discussion on the role of boundary (conditions)
for holography in flat spacetimes can be found in \cite{Marolf:2006bk}.

The relationship with the standard ADM \cite{Arnowitt:1962hi} treatment
at spatial infinity was presented in \cite{Brown:1992br, Brown:2000dz}. The
quasilocal energy agrees with the ADM energy in the limit so that the
spatial boundary is pushed to infinity.

It is also well known that the gravitational action contains divergences even at
tree-level --- they arise from integrating over the infinite volume of the
spacetime. For any variational principle it is possible to add to the action terms
that depend on the fixed boundary data. Thus, the action can be regularized
by supplementing the quasilocal formalism by boundary terms ({\it counterterms})
(see \cite{Lau:1999dp} for asymptotically flat spacetimes\footnote{A rigorous
justification for using these counterterms can be found in \cite{Mann:2005yr}
(see also \cite{Mann:2006bd, Astefanesei:2006zd}).}, \cite{skenderis} for
asymptotically AdS spacetimes, \cite{Cai:1999xg} for a general dilaton potential
and, e.g., \cite{Astefanesei:2008wz} for AdS gravity
with higher derivative terms) that depend on the intrinsic geometry of the regularizing
surface. In this way the difficulties associated with the choice of a reference
background are avoided.

Brown and York have also proposed a quasilocal stress-energy-momentum of
gravitational field which is obtained by varying the action with respect
to the metric on the boundary of the quasilocal region.

A concrete expression
for the {\it regularized} `boundary' stress-energy tensor when the spatial
boundary is pushed to infinity was given in \cite{Astefanesei:2005ad} --- it
was obtained from varying the action supplemented with the counterterms. This
provides not just a concrete method to compute the conserved charges and to
study thermodynamical properties of black objects, but also gives the answer
in a form which is {\it holographic} in spirit.

It is important to emphasize that, despite the similarity between the
definitions of the boundary stress tensor ($\tau^{ij}$) and the standard
matter stress tensor ($T^{\mu\nu}$), they have completely different physical
interpretations \cite{Brown:1992br}.

First of all, it should be noticed
that $\tau^{ij}$ characterizes the {\it entire} system, including
contributions from both the gravitational field and the matter fields.
On the other hand, for a background which satisfies the equations of
motion for gravity and matter, the boundary stress tensor satisfies
an approximate local conservation law \cite{Brown:1992br}
\begin{eqnarray}
\label{eq:conservation}
D_i\tau^{ij}=-T^{nj}
\end{eqnarray}
Here, $D_i$ is the covariant derivative of the induced metric on the boundary 
and $T^{nj}=n_{\mu}T^{\mu j}$ ($n_\mu$ is the normal on the boundary). The source term on the right-hand
side vanishes under the assumptions that matter fields fall off sufficiently
fast at infinity (no matter in a neighborhood of the boundary). Obviously, when
the component of $T^{nj}$ in the direction of a Killing vector $\xi^i$
vanishes, the matter fields do not play any role in defining the conserved
charge associated with this Killing vector.

One would like to understand how this formalism is generalized to unbalanced
solutions --- this is especially interesting in connection with the black ring/blackfold
approach \cite{Emparan:2007wm,Caldarelli:2008pz, Emparan:2009cs}. In the black ring/blackfold approach,
one can construct new black hole solutions in higher dimensions by using
(thin) black branes curved into a specific shape. Thus, an important question is
if the solution remains regular after bending.

We will see that for the unbalanced solutions
there can also exist a source term in the right-hand side of (\ref{eq:conservation})
due to the conical singularities in the metric. Based on this observation, we
will use the conservation of the boundary stress tensor to obtain the equilibrium
condition for a {\it general} (thin or fat) black ring solution (the dynamic
balance condition for a {\it thin} black ring in five dimensions was first observed
in \cite{Elvang:2003mj, Emparan:2004wy}
and obtained in higher dimensions in \cite{Emparan:2007wm}).

In this paper, we also carry out some preliminary investigation of the
thermodynamics of the unbalanced solutions and present a `generalized'
quantum statistical relation. In particular, we use the quantum statistical
relation to read off  the term due to the stresses. In the limit for which
this term vanishes, we will obtain the dynamic balance condition, or, in
other words, a `balanced' ring solution with the usual quantum statistical relation.
This can be regarded as an important check of our proposal that
the equilibrium condition can be obtained from the conservation of
the renormalized boundary stress tensor.

At this point it is worth to comment on the relation with the previous
literature. The fact that the ring is not in equilibrium should be
reflected in a non-conservation of the stress tensor was first pointed out in
\cite{Emparan:2007wm}. The thin black ring is approximated by a
black string (a black 1-brane)\footnote{A ring can be obtained by
bending a {\it boosted} black string \cite{Emparan:2004wy}.} and
so, at large distances, the gravitational field is generated by
a source with a distributional stress-energy tensor which has non-zero
components {\it only} along directions tangent to the worldvolume.
The main observation in \cite{Emparan:2009cs} is that, for a thin
brane, the equations of motion \cite{Carter:2000wv} can be obtained
by demanding the conservation of the stress-energy tensor. Thus,
the absence of the external forces in the equations of motion is
equivalent with the conservation of the stress-energy tensor.

However, our situation is slightly different. First of all, we keep
the conical singularity in the boundary and so the `shape' of the
boundary is changed --- we compute the boundary stress tensor with
respect to this metric.\footnote{In ADM formalism, a computation
of the energy for a background with a boundary which contains a conical
singularity was presented in \cite{Nucamendi:1996ac} --- despite this
similarity, our computations are not related in any way with the ones
presented in this paper.} More importantly, we do not have to use
the thin ring approximation, our computations are done for a general
black ring solution. The non-conservation of the boundary stress tensor in
our case should be related to the modification (the existence of conical
singularities) of the standard asymptotics for flat spacetimes.

Consistent with the point raised above, it is important to emphasize 
that the stress tensor considered in \cite{Emparan:2007wm, Emparan:2009cs} 
is defined in the bulk. To 
make the connection with this work more concrete we note that, in 
principle, one could also consider the quasilocal boundary for a 
foliation with the ring topology hypersurfaces. If such an analysis 
is possible, it should capture the information about the stress of 
the {\it conical disk} for the unbalanced solution.

The structure of the paper is the following: in the next section we 
describe the quasilocal formalism and set the conventions 
for the rest of the paper. In Section 3 we obtain the balance condition 
by using the energy conservation of the boundary stress tensor. In 
Section 4 we propose a generalized quantum statistical relation for 
the unbalanced black ring and check consistency with the proposal 
in the previous section. Finally, we conclude with a discussion of 
our results. Some details about the boundary stress tensor of a boosted 
black string are collected in appendix.
 
%%%%%%%%%%%%%%%%%%%%%%%%%%%%%%%%%%%%%%%%%%%%%%%%%%%%%%%%%%%%%%%%%%%%%%
\section{Quasilocal stress-energy tensor}
%%%%%%%%%%%%%%%%%%%%%%%%%%%%%%%%%%%%%%%%%%%%%%%%%%%%%%%%%%%%%%%%%%%%%%

In this section we present a review of the quasilocal formalism,
set the conventions for the rest of the paper, and comment on
the role of energy conservation in understanding the unbalanced
solutions.

In a very basic sense, gravitational entropy can be regarded as
arising from the Gibbs-Duhem relation applied to the path-integral
formulation of quantum gravity \cite{Gibbons:1976ue}. In the semiclassical
limit this yields a relationship between gravitational entropy and other
relevant thermodynamic quantities, such as mass, angular momentum,
and other conserved charges.

The conserved charges of asymptotically flat black holes are usually
computed by using Hamiltonian methods. The gravitational Hamiltonian
must have well defined functional derivatives and must preserve the
boundary conditions on the fields. Since there is a formal connection
between total energy and time translation, it is natural to expect
that the gravitational mass should be related to the value of the
gravitational Hamiltonian --- this idea is at the basis of defining both,
the ADM and quasilocal, masses.

The quasilocal energy is the value of
Hamiltonian which generates unit magnitude proper-time translations
in a timelike direction orthogonal to spacelike hypersurfaces at some
fixed spatial boundary \cite{Brown:1992br} --- it agrees
with the ADM energy in the limit that the spatial boundary is pushed to infinity.

However, it is important to
point out a key difference between the quasilocal mass and the ADM mass
\cite{Brown:2000dz}. In ADM formalism the hypersurfaces are Cauchy
surfaces and so the data on one slice completely determines the future
evolution of the system. That is not the case for the quasilocal formalism:
the hypersurfaces with which we foliate the spacetime are not Cauchy surfaces.

In this paper, we are interested in 5-dimensional stationary solutions.
We choose the observers that are stationary with respect to the boundary,
i.e., their five-velocity
is perpendicular to the normal. These are the observers that will not
stretch or squash the boundary, which would affect the energy. Therefore,
we define the asymptotically flat spacetimes to have the following fall
off behavior \cite{Tomizawa:2009ua}
\begin{eqnarray}
\label{eq:AF}
ds^2=g_{\mu\nu}dx^{\mu}dx^{\nu}
&\simeq& \left(-1+\frac{8M_{ADM}}{3\pi r^2}+{\cal O}(r^{-3})\right)dt^2- \left(\frac{8J_\phi\sin^2\theta}{\pi r^2}+{\cal O}(r^{-3})\right)dtd\phi\nonumber\\
&&- \left(\frac{8J_\psi\cos^2\theta}{\pi r^2}+{\cal O}(r^{-3})\right)dtd\psi \nonumber\\
&& + \left( 1 + {\cal O}(r^{-1}) \right) \left(dr^2+ r^2\left(d\theta^2 + \sin^2\theta d\phi^2 + \cos^2\theta d\psi^2 \right) \right)
\end{eqnarray}
having the spherical spatial infinity, $S^3_\infty$. We use
Greek indices to denote the bulk coordinates.

The validity of this metric in the asymptotic region could always
be used to define mass and angular momenta, $M_{ADM}$ and $J_a$. We
note that the quasilocal definitions are more powerful because they
do not involve a particular coordinate system.

To define the conserved charges within quasilocal formalism, we use
the divergence-free boundary stress tensor proposed in
\cite{Astefanesei:2005ad}:
\begin{eqnarray}
\label{eq:Tij}
\tau_{ij}\equiv\frac{2}{\sqrt{-h}}\frac{\delta I}{\delta h^{ij}}=
\frac{1}{8\pi G}\Big( K_{ij}-h_{ij}K
-\Psi(\mathcal{R}_{ij}-\mathcal{R}\,h_{ij})-h_{ij}\square\Psi+\Psi_{;ij}\Big) \label{Tik}
\end{eqnarray}
where $\Psi=\frac{c}{\sqrt{\mathcal{R}}}$, $h_{ij}$ is the induced boundary metric, 
and $\mathcal{R}_{ij}$ its Ricci scalar. Note that the constant $c=\sqrt{2},\sqrt{3/2}$ for a
boundary topology $S^2\times R\times R$ or $S^3\times R$, respectively. 

The \textit{renormalized} action $I$ in five dimensions considered in the definition
of this stress tensor, including the counterterms, is of the form
\begin{eqnarray}
\label{eq:action}
I&=& \frac{1}{16\pi
G_5}\int_MR\,\sqrt{-g}\,d^5x+\frac{\epsilon}{8\pi G_5}\int_{\partial
M}(K-c\, \sqrt{\mathcal{R}} )\,\sqrt{-h}\,d^4 x
\end{eqnarray}
where $K$ is the extrinsic curvature of $\partial M$, $\epsilon=+1$ where
$\partial M$ is timelike or $\epsilon=-1$ where it is spacelike, and $h$ is the determinant of the induced
metric on $\partial M$. Also, the counterterm considered here involves the Ricci
scalar $\mathcal{R}$ of the induced metric on the boundary $h_{ij}$.

As we have already mentioned in the Introduction, this stress tensor is covariantly
conserved (with respect to the boundary metric) for vacuum {\it regular}
solutions. When there are also matter fields, the equation (\ref{eq:conservation})
expresses the local conservation of the boundary stress-energy up to the
flow of matter energy-momentum across the boundary.

The boundary metric can be written, at least locally, in ADM-like
form
\begin{eqnarray}
\label{sigma}
 h_{ij}dx^i dx^j=-N^2\,dt^2+\sigma_{a b}\,(dy^a
+N^a\,dt)(dy^b +N^b\,dt)
\end{eqnarray}
where $N$ and $N^a$ are the lapse function and the shift vector
respectively and \{$y^a$\} are the intrinsic coordinates on the
hypersurface $\Sigma$. The boundary conditions for the quasilocal
Hamiltonian include fixation of the boundary spatial metric, lapse
function, and shift vector.

Provided the boundary geometry has an isometry generated by a
Killing vector $\xi ^{i}$, a conserved charge
\begin{eqnarray}\label{eq:charge}
 {\mathfrak Q}_{\xi }=\oint_{\Sigma }d^3y\,
\sqrt{\sigma}n^i\tau_{ij}\,\xi^j
\end{eqnarray}
can be associated with the hypersurface $\Sigma $ (with normal
$n^{i}$). Physically this means that a collection of observers on
the hypersurface whose metric is $h_{ij}$ all observe the same value
of ${\mathfrak Q}_{\xi }$ provided this surface has an isometry
generated by $\xi^{i}$.

The mass and the angular momenta are
\begin{eqnarray}
\label{eq:quasimassandmomenta}
M=\oint_{\Sigma} d^3y \sqrt{\sigma} n^i\,\tau_{ij}\,\xi_t^j\,,\qquad J_{\phi}=\oint_{\Sigma} d^3y \sqrt{\sigma} n^i\,\tau_{ij}\,\xi_{\phi}^j\,,\qquad J_{\psi}=\oint_{\Sigma} d^3y \sqrt{\sigma} n^i\,\tau_{ij}\,\xi_{\psi}^j\,\nonumber
\end{eqnarray}
where the normalized Killing vectors associated with the mass and angular
momenta are $\xi_t=\partial_t$, $\xi_{\phi}=\partial_{\phi}$,
and $\xi_{\psi}=\partial_{\psi}$ respectively. As it was recently explicitly
shown in several concrete 5-dimensional black objects examples \cite{AMRS},
the conserved charges from the quasilocal formalism match the ADM charges exactly.

Armed with this formalism we will obtain the balance condition for a general
black ring solution in the next section. As an warm-up exercise let us discuss Myers-Perry
black hole \cite{Myers:1986un} (with one angular momentum) in Boyer-Lindquist coordinates:
\begin{eqnarray}
\label{eq:Kerr-5D}
ds^2 & = & -dt^2 +\Sigma\,\left(\frac{r^2}{\Delta}\,dr^2 + d\theta^2 \right)+(r^2 + a^2)\,\sin^2\theta \,d\psi^2+r^2\cos^2\theta \,d\phi^2 \nonumber \\
&& +\,\frac{m}{\Sigma}\, \left(dt - a\,\sin^2\theta \,d\psi \right)^2
\end{eqnarray}
where
\begin{eqnarray}\label{eq:sigma}
\Sigma=r^2+a^2 \,\cos^2\theta , \;\;\;\;\;\;
\Delta= (r^2 + a^2)r^2  -m \, r^2
\end{eqnarray}
and $\,m \,$  is a parameter related to the physical
mass of the black hole, while the parameter $\,a \,$ is associated
with its angular momentum.

The metric above
depends only on two of the coordinates:
$0<r<\infty$ and $0<\theta<\pi/2$ and is independent
of time $-\infty < t < \infty$, and the angles $0 < \phi,\psi < 2\pi$.
Asymptotically the metric approaches the flat background
\begin{eqnarray}
\label{eq:flat}
ds^2\simeq-dt^2+dr^2+ r^2(d\theta^2+  \cos^2\theta d\phi^2+ \sin^2 \theta d\psi^2)
\end{eqnarray}
The non-vanishing components of the stress tensor (\ref{eq:Tij}) when $c=\sqrt{3/2}$ are
\begin{eqnarray}
\label{Kerr-5D-Tij}
\tau_{tt}&=&\frac{1}{8\pi G_5}\left(-\frac{3}{2}\,m\, \frac{1}{r^3}-\frac{5}{3} a^2 \frac{\cos 2 \theta}{r^3}  +\mc{O}(1/r^5)\right)\nonumber\\
\tau_{t\psi}\equiv \tau_{\psi t}&=&\frac{1}{8\pi G_5}\left(-2 \,a\, m  \frac{\sin^2 \theta }{r^3}+\mc{O}(1/r^5)\right)\nonumber\\
\tau_{\theta\theta}&=&\frac{1}{8\pi G_5}\left(\frac{2}{3} a^2 \frac{\cos 2 \theta}{r}+\mc{O}(1/r^3)\right)\nonumber\\
\tau_{\phi\phi}&=&\frac{1}{8\pi G_5}\left(\frac{2}{3} a^2 \frac{ (1+2 \cos 2 \theta )\,\cos^2 \theta}{r}+\mc{O}(1/r^3)\right)\nonumber\\
\tau_{\psi\psi}&=&\frac{1}{8\pi G_5}\left(\frac{2}{3} a^2 \frac{(-1+2 \cos 2 \theta )\, \sin^2\theta}{ r}+\mc{O}(1/r^3)\right)
\end{eqnarray}
The stress-energy conservation law can be easily checked. For a generic boundary
metric (\ref{eq:flat}) (with $r$ constant) and a stress tensor with
non-vanishing components as in (\ref{Kerr-5D-Tij}), the only non-trivial term is
\begin{eqnarray}\label{eq:conservationexpanded}
D^{i}\tau_{i\theta}=[\partial_{\theta}\left(\sin\theta \cos\theta \, \tau_{\theta\theta}\right)+\tan^2\theta\,  \tau_{\phi\phi}-\cot^2\theta \, \tau_{\psi\psi}](\cos\theta \sin \theta)^{-1}
\end{eqnarray}
By replacing the explicit values from (\ref{Kerr-5D-Tij}) one can easily check
that $D^{i}\tau_{i\theta}$ vanishes.

This term is also playing an important role in our analysis of
unbalanced solution. We will see in the next section that this
term does not vanish when the boundary metric contains conical
singularities --- the `non-conservation' of the boundary stress
tensor measures the deviation from the standard boundary conditions.
The presence of conical singularities in the boundary metric is
not, however, a drastic change (it is not similar, e.g., with changing
the asymptotics to Anti-de Sitter) and this is why we can still
use the boundary stress tensor (\ref{eq:Tij}) to study this type
of spacetime.

%%%%%%%%%%%%%%%%%%%%%%%%%%%%%%%%%%%%%%%%%%%%%%%%%%%%%%%%%%%%%%%%%%%%%%%%%%%%%%%%%%%%%%%%%
\section{Unbalanced black ring}
%%%%%%%%%%%%%%%%%%%%%%%%%%%%%%%%%%%%%%%%%%%%%%%%%%%%%%%%%%%%%%%%%%%%%%%%%%%%%%%%%%%%%%%%%%

In this section, we explicitly check the conservation of the regularized
stress-energy tensor for the unbalanced black ring. We also show how the
conservation law yields the dynamical equilibrium condition for any, {\it thin}
(black ring solutions with parameters $\nu,\lambda<<1$) or \textit{fat}
(those that are not thin), singly spinning black ring.\footnote{Note that this
 classification is equivalent to the one employed in finding the higher
 dimensional \textit{thin} black rings of \cite{Emparan:2007wm} with $r_0<<R$,
 where $r_0$ is the radius of $S^2$  and $R$ is the radius of $S^1$.
 As shown in \cite{Elvang:2003mj} a redefinition of the 
form $\nu=r_0 \sinh^2 \sigma/R$ and $\lambda=r_0\,\cosh^2\sigma/R$
 gives the relationship between the parameters of the black string and the black ring while changing the
 coordinates $r=-R F(y)/y$, $\cos\theta=x$, $z= R\psi$ and taking the $R\rightarrow\infty$ limit.}

%%%%%%%%%%%%%%%%%%%%%%%%%%%%%%%%%%%%%%%%%%%%%%%%%%%%%%%%%%%%%%%%%%%%%%
\subsection{The model}

Using the conventions in \cite{Elvang:2003mj} we can write a general
line element as
\begin{eqnarray}
\label{eq:generalmetricunbalanced}
  ds^2 &=& -\frac{F(x)}{F(y)} \left(dt+
     R\sqrt{\lambda\,\nu} \,(1 + y)\, d\psi\right)^2  \label{ring0}\\
  &&
   +\frac{R^2}{(x-y)^2}
   \left[ -F(x) \left( G(y)\, d\psi^2 +
   \frac{F(y)}{G(y)}\, dy^2 \right)
   + F(y)^2 \left( \frac{dx^2}{G(x)}
   + \frac{G(x)}{F(x)}\,d\phi^2\right)\right]  \nonumber
\end{eqnarray}
with
\begin{eqnarray}
  F(\xi) = 1 - \lambda\xi \, ,
\qquad  G(\xi) = (1 - \xi^2)(1-\nu \xi) \,
\end{eqnarray}
$R,\lambda$ and $\nu$ are parameters whose appropriate combinations
give the mass and angular momentum. The parameters $\nu$ and $\lambda$
have the range $0\leq\nu<\lambda<1$. Asymptotic spatial infinity is
reached as $x\to y\to-1$.

This is a vacuum solution of Einstein equations in five dimensions. To
obtain a Lorentzian signature, the variables $x$ and $y$ are
constraint to take values in
\begin{eqnarray}\label{eq:xyrange}
-1\leq x\leq 1\,,\qquad  -\infty<y\leq-1\,,\quad \lambda^{-1}<y<\infty
\end{eqnarray}
Note that for $y\in [\lambda^{-1},\nu^{-1}]$ the coordinate $t$ becomes
spacelike and so $t$ and $y$ are interchanged.

It is also important to emphasize that, since there are conical
singularities, this is not a {\it regular} solution. Once the conical
singularities are eliminated, we get a physical regular solution (black hole
or black ring).

As shown in \cite{Elvang:2003mj}, in order to balance forces in the
ring one must identify
$\psi$ and $\phi$ with equal periods
\begin{eqnarray}\label{eq:phsiperiod}
\Delta\phi=\Delta\psi=\frac{4\pi \sqrt{F(-1)}}{|G'(-1)|}=
\frac{2\pi\sqrt{1+\lambda}}{1+\nu}
\end{eqnarray}
This eliminates the conical singularities at the fixed-point sets $y=-1$
and $x=-1$ of the Killing vectors $\partial_\psi$ and $\partial_\phi$,
respectively.

However there still is the possibility of conical singularities at $x=+1$.
These can be avoided in either of two ways. Fixing
\begin{eqnarray}\label{eq:lambring}
\lambda=\lambda_c\equiv \frac{2\nu}{1+\nu^2}\qquad {\rm (black\; ring)}
\end{eqnarray}
makes the circular orbits of $\partial_\phi$ close off smoothly also at
$x=+1$. Then $(x,\phi)$ parametrize a two-sphere, $\psi$ parametrizes a
circle, and the solution
describes a black ring. Alternatively, if we set
\begin{eqnarray}\label{eq:lambhole}
\lambda=1\qquad {\rm (black\; hole)}
\end{eqnarray}
then the orbits of $\partial_\phi$ do not close at $x=+1$. Then
$(x,\phi,\psi)$ parametrize an $S^3$ at constant $t,y$. The solution is
the same as the spherical black hole of \cite{Myers:1986un} with a single rotation
parameter.

Both for black holes and black rings, $|y|=\infty$ is an
ergosurface, $y=1/\nu$ is the event horizon, and the inner, spacelike
singularity is reached as $y\to\lambda^{-1}$ from above.

%%%%%%%%%%%%%%%%%%%%%%%%%%%%%%%%%%%%%%%%%%%%%%%%%%%%%%%%%%%%%%%%%%%%%
\subsection{Balance condition}
%%%%%%%%%%%%%%%%%%%%%%%%%%%%%%%%%%%%%%%%%%%%%%%%%%%%%%%%%%%%%%%%%%%%%

The equilibrium condition is a constraint on the
parameters of the unbalanced ring solution \cite{Elvang:2003mj} which
is equivalent with the removing of all conical singularities in the metric.

A nice physical interpretation was given in \cite{Emparan:2004wy}: the absence
of conical singularities is equivalent with the equilibrium of the forces acting
on the ring. A black ring can be obtained by bending a {\it boosted} black string.
Thus, the linear velocity along the string becomes the angular velocity of the
black ring. The equilibrium of centrifugal and gravitational forces imposes
a constraint on the radius of the ring $R$, the mass, and the angular momentum. In
this way one can see that, indeed, just two parameters are independent in the
solution of the neutral black ring. A discussion of black string within
the quasilocal formalism is presented in the following section.

The previous observation was used in \cite{Elvang:2003mj, Emparan:2004wy} and \cite{Emparan:2007wm} to
obtain a balanced condition for the {\it thin} rings. By using the conservation
of the boundary stress tensor (\ref{eq:Tij}), we obtain the balance condition
for a general ({\it thin} or {\it fat}) black ring solution.

The quasilocal formalism is employed in the asymptotic region and so only
conical singularities in the boundary can be detected in the boundary stress
tensor. Therefore, one should choose a general enough foliation so that the
induced metric of the boundary contains this non-trivial information.

Unlike in the previous literature where the conical singularity in the boundary
$x=y=-1$ was eliminated first, we will just get rid of the conical singularity
in the bulk. Thus, to avoid the conical singularity in
the bulk, one must identify $\psi$ and $\phi$ with an equal period
\begin{eqnarray}
\Delta\phi=\Delta\psi=\frac{4\pi \sqrt{F(+1)}}{|G'(+1)|}=
2\pi\frac{\sqrt{1-\lambda}}{1-\nu}
\label{eq:bulkbalance}
\end{eqnarray}
For the reasons mentioned in Section 2 and since the computations simplify,
we prefer to change the coordinates as follows:
\begin{eqnarray}
\label{eq:coordenas}
x\to -1+\frac{2 \,m \cos^2\theta}{ \left(r^2+a^2 \cos^2\theta \right)}\,,\qquad y\to -1-\frac{2 \,m\sin^2\theta }{ \left(r^2- m+a^2 \cos^2\theta\right)}
\end{eqnarray}
In order to get an asymptotic metric of the form (\ref{eq:flat}), we work in
the following gauge:
\begin{eqnarray}
\label{eq:relabelingcarges}
m= \frac{(1+\lambda )^2 R^2 }{1+\nu }\,,\qquad a= \frac{(1+\lambda )^{1/2}  R(\lambda -1+\nu +3 \lambda  \nu )^{\frac{1}{2}}}{(1+\nu )}
\end{eqnarray}
To gain some intuition about these coordinates, let us consider a $5$-dimensional
spinning black hole with one angular momentum --- it corresponds to $\lambda=1$ in
(\ref{eq:generalmetricunbalanced}). By the coordinate transformation (\ref{eq:coordenas})
the solution takes exactly the form of (\ref{eq:Kerr-5D}) with the
additional identification of parameters,
\begin{eqnarray}
m=\frac{4\,R^2}{(1+\nu)}\,,\qquad a=\frac{2^{3/2}\,R\sqrt{\nu}}{(1+\nu)}
\end{eqnarray}
Now, let us consider the unbalanced solution (\ref{eq:generalmetricunbalanced}) with the
conical singularity in the bulk removed by (\ref{eq:bulkbalance}). In the coordinates
(\ref{eq:coordenas}) the asymptotic form of the metric is
\begin{eqnarray}
\label{eq:asymptmetric}
g_{tt}&=&-1+\frac{2 R^2 \lambda  (1+\lambda )}{(1+\nu) }\frac{1}{r^2}+\mc{O}(1/r^4)\\
g_{t\psi}&=&\frac{2 R^3 (1+\lambda )^2 \sqrt{\lambda  \nu\,(1-\lambda) }  }{(1-\nu^2) }\frac{\sin^2\theta}{r^2}+\mc{O}(1/r^3)\\
g_{rr}&=&1+\mc{O}(1/r^2)\,\\
g_{\theta\theta}&=&r^2+\mc{O}(1)\\
g_{r\theta}&=&\mc{O}(1/r^3)\,\\
g_{\phi\phi}&=&\frac{ (1-\lambda)(1+\nu )^2 }{(1-\nu)^2(1+\lambda ) }\,r^2 \cos^2\theta +\mc{O}(1)\\
g_{\psi\psi}&=&\frac{ (1-\lambda)(1+\nu )^2 }{(1-\nu)^2(1+\lambda )}\,r^2 \sin^2\theta +\mc{O}(1)
\end{eqnarray}
and the non-vanishing stress tensor components are
\begin{eqnarray}\label{eq:tBR}
\tau_{tt}&=&\frac{1}{8 \pi G_5}\left(-\frac{R^2 (1+\lambda ) (9 \lambda  (1+\nu )-10 (-1+\lambda -2 \nu ) \cos 2 \theta )}{3 r^3 (1+\nu )^2}+\mc{O}(1/r^4)\right)\\
\tau_{\psi t}&=&\frac{1}{8 \pi G_5}\left(-\frac{4 R^3 (1+\lambda )^{5/2} \sqrt{\lambda  \nu }}{(1 +\nu )^2}\frac{\sin^2\theta}{r^3}+\mc{O}(1/r^4)\right)\\
\tau_{\theta\theta}&=&\frac{1}{8 \pi G_5}\left(\frac{4 R^2 (1+\lambda ) (1-\lambda +2 \nu ) \cos 2 \theta }{3 r (1+\nu )^2}+\mc{O}(1/r^3)\right)\\
\tau_{\phi\phi}&=&\frac{1}{8 \pi G_5}\left(\frac{4  R^2 (1-\lambda)(1-\lambda +2 \nu ) \cos^2\theta  (1+2 \cos 2 \theta )}{3 r(1-\nu)^2}+\mc{O}(1/r^3)\right)\\
\tau_{\psi\psi}&=&\frac{1}{8 \pi G_5}\left(\frac{4 R^2 (1-\lambda)  (1-\lambda +2 \nu ) (-1+2 \cos 2 \theta ) \sin^2\theta }{3 r (1-\nu)^2}+\mc{O}(1/r^3)\right)
 \end{eqnarray}
We find that, due to the existence of conical singularities, this
stress tensor is not covariantly conserved with
respect to the boundary metric (\ref{eq:flat}) (with $r$ constant).
In other words, the fact that the stress tensor is not conserved is
reflected in the existence of additional stresses that deform (to some
extent) the boundary metric.

We observe that the stress tensor is only conserved (is equivalent with
the absence of external forces) when
 \begin{eqnarray}
\lambda=\frac{2\nu}{1+\nu^2}
 \end{eqnarray}
which is the dynamic equilibrium condition for black ring. We
expect that this will also be confirmed for other \textit{unbalanced}
solutions such as the doubly spinning black ring \cite{Morisawa:2007di}
multi black hole solutions \cite{Elvang:2007rd,Iguchi:2007is,Elvang:2007hs}
and the new black holes of \cite{Lu:2008js}.

%%%%%%%%%%%%%%%%%%%%%%%%%%%%%%%%%%%%%%%%%%%%%%%%%%%%%%%%%%%%%%%%%%%%%%
\section{Stresses and quantum statistical relation}
%%%%%%%%%%%%%%%%%%%%%%%%%%%%%%%%%%%%%%%%%%%%%%%%%%%%%%%%%%%%%%%%%%%%%%

In this section we present a `generalized' quantum statistical relation
for the unbalanced black ring. We also explicitly obtain the dynamic
equilibrium condition for the black ring in the limit for which the term
due to the stresses vanishes.

For completeness, we start by
presenting a brief review of the gravitational tension
and discussing the black string solution (including the boundary conditions
that permit variations of its length) (see e.g. \cite{Obers:2008pj} and
references therein). In the second part of this section we compute the
action of the unbalanced black ring and show that it satisfies a generalized
quantum statistical relation.

%%%%%%%%%%%%%%%%%%%%%%%%%%%%%%%%%%%%%%%%%%%%%%%%%%%%%%%%%%%%%%%%%%%%%%
\subsection{Gravitational tension}

We start by examining a boosted black string with a {\it fixed}
length $L$ --- the stress tensor components are presented in the Appendix. We
obtain
\begin{eqnarray}
\label{eq:limit}
\lim_{r \to \infty}\sqrt{-h}\left(c\sqrt{\,\mathcal{R}}-K\right)&=&\frac{ r_0 }{2} \sin\theta+\mc{O}(1/r)\nonumber
\end{eqnarray}
where $r_0$ is the event horizon and $\sigma$ the boost parameter in the black string solution (\ref{eq:metricBS}),
so the total action for the black string is
\begin{eqnarray}\label{eq:BSaction}
I\equiv\beta \,G=\beta \frac{r_0}{4G_5}\,L
\end{eqnarray}
where $\beta=1/T$ is the inverse of the temperature.

The other thermodynamical quantities, the linear mass density, the temperature,
and the event horizon area per unit length are
\begin{eqnarray}\label{eq:chargesBS}
\frac{M}{L}= \frac{r_0}{4 G_5}(1+\cosh^2\sigma )\,,\qquad T=\frac{(\cosh\, \sigma)^{-1}}{4\pi r_0} \,,\qquad \frac{\mathcal{A}}{L} = 4\pi r_0^2 \cosh \sigma\,
\end{eqnarray}
The quantum statistical relation contains an additional term
\begin{eqnarray}\label{eq:gdrelationBS}
G-(M-T\, S)=-\frac{r_0\,L\, \sinh^2\sigma }{4 G_5}\,
\end{eqnarray}
This non-trivial term can be interpreted as a new term, $v\,p$, where the
linear momentum $p$ and the boost velocity $v$ are
\begin{eqnarray}\label{eq:pandv}
p=\frac{r_0 \,L}{4 G_5}\cosh\sigma \sinh\sigma\,,\qquad v=\tanh\sigma 
\end{eqnarray}
The boost velocity appears as an intensive parameter in the first law.

It is important to emphasize again that by bending a string to form a circle
one can obtain a black ring with the horizon topology $S^1\times S^2$. One way
to keep this configuration in equilibrium is to add angular momentum such that
the repulsive centrifugal force balances the tension and gravitational
self-attraction.

Since the boost velocity becomes the angular velocity after
bending, one could intuitively guess that the equilibrium is obtained for some
specific values of the boost parameter. Indeed, as observed in \cite{Elvang:2003mj,
Emparan:2004wy} the equilibrium condition for a thin black ring is equivalent
with demanding a vanishing tension for the black string --- in analogy with
the ADM definition, within the quasilocal formalism the tension is
proportional to the $\tau_{zz}$ component of boundary stress tensor. Using
the value of $\tau_{zz}$ from the Appendix, we observe that only
the tensionless black string with a boost parameter
$\sinh^2{\sigma}=1$ is a black ring in the thin limit.

The tension of a spacetime arises as an extension of the usual ADM gravitational
charges when there are additional parameters that characterize the spacetime at
infinity \cite{Traschen:2001pb}. In this case, one can obtain two solutions with
slightly shifted values of the parameters but with the same asymptotics
\cite{Kastor:2006ti,Kastor:2007wr,Harmark:2003eg}.

A general definition for gravitational tension was given in \cite{Harmark:2004ch} --- however, as
in the case of gravitational energy, it can only be defined with respect to a
reference background. A definition of gravitational tension by using the renormalized
boundary stress tensor is clearly more advantageous since the difficulties associated
with the choice of a reference background are avoided.

Let us also briefly comment on the case of a static black string with the boundary conditions
that permit variations of its length \cite{Kastor:2006ti}. In this case, there is an
additional `work' term in the first law given by the product of the
tension and the variation of the length at spatial infinity \cite{Traschen:2001pb}.

Since the computation of the boundary stress tensor for the static black
string is similar with the one in Apendix, we do not present the results here.
However, as expected, our results match the results in \cite{Kastor:2006ti} and the
quantum statistical relation contains a new term due to the tension.

More importantly, it was shown in \cite{Kastor:2006ti} that, when the solutions
are characterized by more than one modulus, the result can be simply
stated in analogy with the physics of elastic materials \cite{Landau}:
the role of stresses is played by a tension tensor. Within the quasilocal
formalism, the strain tensor should encode the changes in the boundary
metric: in shape (e.g., when there exist conical singularities) and/or in size.

%%%%%%%%%%%%%%%%%%%%%%%%%%%%%%%%%%%%%%%%%%%%%%%%%%%%%%%%%%%%%%%%%%%%%%
\subsection{Generalized quantum statistical relation}

We have shown in Section 3 that the conservation of the
boundary stress tensor provides the required dynamical balance condition
for the solution (\ref{eq:generalmetricunbalanced}). However, the quasilocal
formalism also provides a definition for the action which is related to
the thermodynamical potential $G$. Thus, it is important to understand how
the standard thermodynamics is changed for the unbalanced
black ring solution.

To compute the {\it renormalized} action (\ref{eq:action}) we observe
that the scalar curvature $R$ vanishes and so the only contributions
are due to the surface terms. We obtain the following grand-canonical potential:
\begin{eqnarray}\label{eq:gibbs}
G\equiv\frac{I}{\beta}=\beta\frac{k^2 R^2 }{16\pi G_5 }\frac{ \lambda(1-\lambda )  (1+\nu )}{ (1-\nu )^2}
\end{eqnarray}

We also compute the mass and the angular momentum
\begin{eqnarray}\label{eq:massandmomBR}
M=\frac{3 k^2 R^2 }{16 G_5 \pi }\frac{\lambda (1 + \lambda)}{(1 + \nu)}\,,\qquad J_{\psi}=\frac{k^2 R^3}{8 G_5 \pi  }\frac{ (1+\lambda )^{5/2} \sqrt{\lambda  \nu }}{(1+\nu )^2}
\end{eqnarray}
where $k$ is the periodicity of $\phi$ and $\psi$, namely $\Delta\phi=\Delta\psi=k$.
The event horizon, the temperature, and the angular velocity are
\begin{eqnarray}\label{eq:areaandvelBR}
\mathcal{A} =\frac{2 k^2 R^3 (1-\lambda ) \sqrt{\lambda } (\lambda -\nu )^{3/2}}{(1-\nu )^2 (1-\nu )}\,,\,\,\, T=\frac{1}{4\pi R}\frac{1-\nu }{\sqrt{\lambda\,(\lambda -\nu )}}\,,\,\,\,\Omega=\frac{1}{R}\frac{(1-\nu ) }{ \lambda (1 + \nu )}\sqrt{\frac{\lambda \, \nu}{1-\lambda}}
\end{eqnarray}
Note that since we removed the bulk conical singularity and kept the one in
the boundary metric, the charges differ from the ones in
the original paper \cite{Elvang:2003mj}. Of course, as the balance condition
is imposed both results for the charges agree.

The balanced black ring solutions satisfy the standard quantum statistical relation
 \begin{eqnarray}
G=M- T S-\Omega J_{\psi}
 \end{eqnarray}
In our case, the solution is not balanced and we obtain the following generalized
quantum statistical relation:
 \begin{eqnarray}
G-(M- T S-\Omega J_{\psi})=\frac{k^2 R^2}{16 G_5 \pi } \,f[\nu,\lambda]
 \end{eqnarray}
It seems that the extra term, $f[\nu,\lambda]$, should correspond to the stresses
due to the conical singularity in the boundary. This term can be written as
 \begin{small}
 \begin{eqnarray}\label{eq:fnulambda}
f[\nu,\lambda]=\left(\frac{ (1-\lambda ) (3 \lambda+ (\lambda-2)\nu)}{(1-\nu )^2}+\frac{2 (1+\lambda )^{5/2} (1-\nu ) \nu }{\sqrt{1-\lambda } (1+\nu )^3}-\frac{3 \lambda  (1+\lambda )}{1+\nu }\right)
 \end{eqnarray}
 \end{small}
It is zero for $\lambda=2\nu/(1+\nu^2)$ which corresponds to the
equilibrium condition for the asymptotically flat black
ring.\footnote{Note that all the other roots of $f[\nu,\lambda]=0$ that
satisfy the condition $0\leq\nu<\lambda< 1$ do not correspond to
(standard) asymptotically flat metrics.}

%%%%%%%%%%%%%%%%%%%%%%%%%%%%%%%%%%%%%%%%%%%%%%%%%%%%%%%%%%%%%%%%%%%%%%
\section{Discussion}
%%%%%%%%%%%%%%%%%%%%%%%%%%%%%%%%%%%%%%%%%%%%%%%%%%%%%%%%%%%%%%%%%%%%%%
In this work, we used the conservation of the {\it renormalized}
boundary stress-energy  tensor \cite{Astefanesei:2005ad} to obtain
the equilibrium condition for a {\it general} (thin or fat) black
ring solution. This closes a gap left unanswered in the previous
literature that dealt just with the thin ring.

We have also investigated the role of stresses in the thermodynamics of
the unbalanced solutions and proposed a generalized quantum statistical
relation for the unbalanced black ring.

The role of the spatial stress was already pointed out by
Brown and York \cite{Brown:1992br} where a discussion on the thermodynamics of static black
holes with respect to a boundary at finite $r=R=constant$ (in the bulk)
was presented. They provided an interpretation for the trace of spatial
stress as a surface pressure. A straightforward generalization of this
definition to (general) unbalanced solutions when the boundary is at
infinity seems unlikely and we leave a detailed analysis of this
non-trivial issue for future work.

The main goal of this work was to understand how the proposal of
\cite{Emparan:2007wm}, that the dynamical balance of a {\it thin} black ring is related
to the conservation of the stress tensor, can be extended to a general
black ring solution. However, to reach this goal, we have used a slightly
different method, namely the quasilocal formalism supplemented with the
counterterms. The advantage of this method is that the problems associated
with the background subtraction are avoided. Also, one can compute the
action (on the Euclidean section) not just the conserved charges and so
this method provides a complete way to study the
thermodynamics.

A concrete example that was extensively studied in the literature was a
black string with boundary conditions that permit variations of its length.
The first law in this case can be expressed as
\begin{equation}
dM=TdS+\varGamma dL
\end{equation}
where $\varGamma$ is the tension (it is proportional to
$\tau_{zz}$ component of the spatial stress) and $L$ is the length
of the string.\footnote{Since the mass is proportional to $L$,
at first sight, the new term in the first law seems problematic.
However, one can see that for holding a finite horizon area the
mass parameter (that is the monopole in a multipoles expansion of
$g_{tt}$ component of the metric) must be varied in a precise way
as $L$ is varied. Thus, to get the correct first law one should
express first the mass in terms of the horizon area and the length
$L$ as independent variables \cite{Kastor:2006ti}.}

It seems that the only component of the spatial stress which plays a role
in thermodynamics is $\tau_{zz}$, where $z$ is the direction along the string.
Intuitively, one can easily understand that this component of the stress
tensor is related to the tension of the string. However, it is not so
obvious why the other components of the spatial stress do not play any role.
In fact, one expects that when the boundary is at a finite distance in the
bulk the analysis of Brown and York should also apply to black string and
so all the components of the spatial stress should play a role. However, when the
boundary is pushed to infinity the only relevant contribution is coming from
the $\tau_{zz}$ component.

A connection between black strings and (thin) black rings was pointed out
in \cite{Elvang:2003mj, Emparan:2004wy}. That is the equilibrium condition
for the (thin) black ring is
equivalent to a vanishing tension of the black string. In other words,
the only black strings that are obtained  as a limit from black rings are the
{\it tensionless} ones. This fact was further explored in \cite{Emparan:2007wm,
Emparan:2009cs} and it
was found that this connection is more subtle: the equilibrium condition can
be obtained from the conservation of the stress-energy tensor.
We have explicitly checked that the equilibrium condition for
a general black ring solution can be obtained from the conservation of
the renormalized boundary stress tensor.

By providing a generalized
quantum statistical relation for the unbalanced black ring we also made
a step in understanding the thermodynamics of unbalanced solutions.
We have reached a similar conclusion as in \cite{Kastor:2006ti} for extended objects.
In our case the strain tensor encodes the modifications in the shape and
size of the boundary metric. It is important to note, though, that our
proposal can be used not just for extended objects (e.g., black strings),
but also for more general (unbalanced) solutions.

As a final comment, we note that the quasilocal formalism supplemented with
counterterms is a very robust method to study the thermodynamics of black objects
and it may be useful in understanding the holography in flat space.

\acknowledgments

We would like to thank Oscar Varela for useful conversations. DA would 
also like to thank Niels Obers for interesting discussions. This work 
was supported by the German-Israeli Project cooperation (DIP H.52) and 
the German-Israeli Fund (GIF).

%%%%%%%%%%%%%%%%%%%%%%%%%%%%%%%%%%%%%%%%%%%%%%%%%%%%%%%%%%%%%%%%%%%%%%
\appendix
\section{Stress tensor for boosted black string}
%%%%%%%%%%%%%%%%%%%%%%%%%%%%%%%%%%%%%%%%%%%%%%%%%%%%%%%%%%%%%%%%%%%%%%

A simpler perhaps but more intuitive example to understand the boundary
stress tensor is the boosted black string

\begin{eqnarray}
 \label{eq:metricBS}
ds^2=-\hat{f}\left(dt-\frac{r_0\sinh\sigma\cosh\sigma}{r\,\hat{f}} dz\right)^2+\frac{f}{\hat{f}}\,dz^2+\frac{dr^2}{f}+r^2 (d\theta^2+ \sin^2 \theta\, d\phi^2)
\end{eqnarray}
where
\begin{eqnarray}
f=1-\frac{r_0}{r},\qquad\hat{f}=1-\frac{r_0 \cosh^2\sigma}{r}
\end{eqnarray}

The non-trivial components of the boundary stress tensor for a black
string with a boost (parametrized by $\sigma$) in five dimensions are
\begin{eqnarray}
\label{string-5D-Tik}
\tau_{tt}&=&\frac{1}{8 \pi G_5}\left(-\frac{r_0}{2} (1+ \cosh^2{ \sigma})\,\frac{1}{ r^2} +\mc{O}(1/r^3)\right)\,,\nonumber\\
\tau_{tz}&=&\frac{1}{8 \pi G_5}\left(-\frac{r_0}{2}  \cosh \sigma  \sinh\sigma \,\frac{1}{r^2}+\mc{O}(1/r^3)\right)\,,\\
\tau_{\theta\theta}&=&\frac{1}{8 \pi G_5}\left(-\frac{5}{8}\frac{r_0^2}{r}+\mc{O}(1/r^2)\right)\,,\nonumber\\
\tau_{\phi\phi}&=&\frac{1}{8\pi G_5}\left(-\frac{5}{8} \frac{r_0^2}{r}\, \sin^2\theta+\mc{O}(1/r^2)\right)\,,\nonumber\\
\tau_{zz}&=&\frac{1}{8 \pi G_5}\left(\frac{r_0}{2} (1-\sinh^2{\sigma})\,\frac{1}{r^2}+\mc{O}(1/r^3)\right)\nonumber
\end{eqnarray}
The boundary stress tensor also satisfies the conservation law $D^{i}\tau_{ij}=0$. In this case the covariant derivative is with respect to the following asymptotic
black string metric (instead of (\ref{eq:flat}) for black ring)
\begin{eqnarray}
\label{eq:flatBS}
ds^2\simeq-dt^2+dr^2+ r^2(d\theta^2+ \sin^2 \theta\, d\phi^2)+dz^2
\end{eqnarray}
and the only non-trivial covariant derivative is
\begin{eqnarray}\label{eq:BSconsvlaw}
D^{i}\tau_{i\theta}=[\sin\theta\,\partial _{\theta }\, (\sin \theta \,\tau_{\theta\theta})-\cot\theta \,\tau_{\phi\phi}](\sin \theta)^{-2}
\end{eqnarray}
Replacing (\ref{string-5D-Tik}) in the latter equation it is possible to
check that the boundary stress tensor is indeed covariantly conserved for
any value of the parameters in the solution.

%%%%%%%%%%%%%%%%%%%%%%%%%%%%%%%%%%%%%%%%%%%%%%%%%%%%%%%%%%%%%%%%%%%%%%

\end{document}